# Giant and negative dielectric tunability induced by interfacial polarization in $Pb(Fe_{1/2}Nb_{1/2})_{1-x}Ti_xO_3$ single crystals

Kui Liu,[1] Xinyi Zhang,[1,*] and Jingzhong Xiao[1,2]

[1]*Synchrotron Radiation Research Center, Physics Department, Surface Physics Laboratory (State Key Laboratory) of Fudan University, Shanghai, 200433, China*

[2]*International Centre for Materials Physics, Chinese Academy of Sciences, Shenyang 110016, China.*

The giant and negative dielectric tunability of $Pb(Fe_{1/2}Nb_{1/2})_{1-x}Ti_xO_3$ single crystals is reported. A low field of 120 V/cm can induce a great reduction of the capacitance, and the tunability is larger than 80% in low frequency range (<1 MHz) at room temperature. This giant tunability is ascribed to the interfacial polarization at the interface of electrode/sample. A negative dielectric tunability detected only in the tetragonal sample can be also attributed to the interfacial polarization. The origin of the giant and negative tunabilities is discussed with the multipolarization-mechanism model and equivalent circuit model, respectively.

**Introduction**

In recent years, great attention has been paid to the materials with high dielectric tunability due to their potential applications in many electrically tunable microwave devices, such as voltage-controlled oscillators, band pass filters, and phase shifters.[1-3] Many ferroelectric materials were reported to exhibit the high tunability property. Tanmoy *et al* reported that a large tunability as high as 93% was achieved in $BaZr_xTi_{1-x}O_3$ ceramics.[4] M. Jain *et al* studied $Pb_xSr_{1-x}TiO_3$ thin films whose tunability was as high as 70%.[5] However, the high tunability usually needs very high electric field, about 10~100 kV/cm. Therefore, the samples are mostly thin films. Otherwise, high voltage is required for bulk samples. Very recently, Chang-Hui Li *et al* reported a giant dielectric tunability induced by a low electric field (10~100 V/cm) in bulk $LuFe_2O_4$.[6] They suggested that Schottky barrier effect might play a role. Moreover, C. C. Wang *et al* and Guo-Zhen Liu *et al* observed an enhanced tunability due to interfacial polarization.[7-8] The interfacial polarization associated with Maxwell-Wagner relaxation usually occurs at the interface of inhomogeneous regions with different conductivities. A depletion layer is formed at the interface, and a small voltage may induce a considerable dielectric reduction.[9]

Lead iron niobate $Pb(Fe_{1/2}Nb_{1/2})O_3$ (PFN) belongs to the lead based perovskite family exhibiting high dielectric constant and multiferroic properties, and is of great interest for high dielectric constant multilayer capacitors and multimemory devices.[10-11] $Pb(Fe_{1/2}Nb_{1/2})_{1-x}Ti_xO_3$ (PFNT) which is a modified multiferroic material based on PFN has been also studied intensively in recent years.[12-13] However, to our knowledge, the dielectric tunability property of PFNT single crystals has not been reported in the literature up to now. In this letter, we investigated the tunability property of PFNT single crystals at low electric field and low frequency range. Giant and negative tunability was observed, which we suggested to be induced by interfacial polarization.

**Experiment**

High quality PFNT single crystals (x=0.07, named sample A; x=0.48, named sample B) were fabricated without impure phases. The details of the single crystal growth can be found elsewhere.[14] Both samples were cut along plane parallel to (100) face. The thicknesses of sample A and sample B are 0.74 and 0.69 mm, respectively. The powder X-ray diffraction patterns were measured for the phase and plane identification by a Philips X-ray diffractometer with Cu *Kα* line. The XRD results show that the samples are pure perovskite structure. The Raman spectra recorded at room temperature reveal that the sample A is in rhombohedral phase while sample B is in tetragonal phase. The rhombohedral and tetragonal phases are most common non-cubic variants of perovskite structure. Frequency and voltage dependences of dielectric spectra were measured by a HP4284 impedance analyzer with a computer controlled system at room temperature. The frequency range used was from 100 Hz to 1 MHz, and the voltage range varied from 0 to 10 V. The probe of an AC electric field was confined to be 10 mV. The gold and aluminium electrodes were used to check the interface effect, respectively. The Au electrodes were sputtered, and Al electrodes were evaporated on both sides of samples, respectively.

**Results and discussion**

### A. Giant dielectric tunability

Figure 1 shows the electric field dependence of the normalized capacitance ($N_c$-$E$ curves), $N_c=C_p(E)/C_p(E=0)$, for sample A and sample B, respectively, at various frequencies (shown in the figure). At low frequencies (for example, 10 kHz for sample A and 1 kHz for sample B), the normalized capacitance exhibits a rapid drop at low electric field range followed by a slow decrease at high fields. As the frequency is increased, the rapid drop becomes weaker, and then disappears with only slow decrease left when the frequency is higher than 100 kHz for sample A and 10 kHz for sample B, respectively. At 1MHz, the normalized capacitance of sample B is nearly unvaried with electric field. The similar behavior was observed in $La_{0.7}Sr_{0.3}MnO_3/BaTiO_3$ multilayer films, and the enhanced tunability was attributed to the interfacial polarization.[7] The relative tunability ($n_r$) is defined by[4,6]

$$n_r = \frac{C_p(E=0) - C_p(E)}{C_p(E=0)} \times 100\%. \qquad (1)$$

According to Eq. (1), we can calculate the maximum tunability, which is obtained to be as high as 85% for sample A and 94% for sample B. It is worth notice that the giant tunability is induced by a much lower electric field (about 100 V/cm) in our PFNT single crystal samples.

Figure 2 shows the complex impedance spectra (Z' vs. Z'') at four selected voltages for both samples. The inset is the frequency dependence of capacitance at selected voltages. As shown in Figs. 2(a) and 2(b), two semicircular arcs in impedance spectra were detected in both samples, corresponding to two electrical responses of impedance. For sample A, one semicircular arc is in the low frequency range, $f<3.0 \times 10^5$ Hz, the other one is in the high frequency range, $f>3.0 \times 10^5$ Hz. For sample B, the low and high frequency responses are in the range $f<5.5 \times 10^4$ Hz and $f>5.5 \times 10^4$ Hz, respectively. For both samples, the low frequency response is strongly dependent on voltage, while the high frequency response depends weakly on voltage dependence. The impedance spectra suggest that the interfacial polarization has a considerable effect on the dielectric property. Moreover, as seen from the insets of Figs. 2(a) and 2(b), a step decrease appears in the capacitance-frequency curves, which is also a typical characteristic of Maxwell-Wagner relaxation, indicating that the interfacial polarization dominates in the low frequency range. Therefore, we can conclude that the giant tunability is caused by the interfacial polarization.

Figure 3 shows the electric field dependence of normalized capacitance of sample A (at 100

kHz) and sample B (at 10 kHz) when the Al electrodes were used. The tunability was greatly suppressed when Al electrodes were used. For sample A, the tunability is dropped to 5% at $f$=100 kHz and $E$=120 V/cm, much smaller than 48% for Au electrodes at the same condition. For sample B, its tunability is only 2% at $f$=10 kHz and $E$=120V/cm, which is also much smaller than 55% in case that Au electrodes were used. Since the tunability of both samples is significantly dependent on the electrode materials, we suggest that the interfacial polarization might be mainly caused by the electrode effect. The physical mechanism can be explained as follows. There are some point defects such as oxygen vacancies and $Fe^{2+}$ ions in our samples due to oxygen deficiency during the crystal growth. Therefore, our crystals exhibit semiconductor-like behavior and depletion layers are formed at the interface of electrode/sample. At high frequencies ($f$>$10^6$ Hz for sample A; $f$>$10^5$ Hz for sample B), the charge carriers can not move far away from the defect center, so the potential barrier of the depletion layer has no effect on the conductivity. In this case, the bulk contribution dominates. While at low frequencies ($f$<$10^4$ Hz for sample A; $f$<$10^3$ Hz for sample B), the charge carriers have enough time to move far away from the defect center, and the conductivity might be blocked by the potential barrier of the depletion layer. Then the charge carriers pile up at the interface of electrode/sample, leading to the interfacial polarization associated with Maxwell-Wagner relaxation. In this case, the interfacial contribution dominates.

A multipolarization-mechanism model proposed by Chen Ang *et al* can be used to quantitatively analyze the interfacial and bulk contributions to the dielectric tunability.[15] The normalized capacitance based on the multipolarization-mechanism model is described as

$$N_c = \frac{C_p(E)}{C_p(E=0)} = x[\cosh aE]^{-2} + (1-x)\frac{1}{(1+bE^2)^{1/3}}, \qquad (2)$$

where, $a$ and $b$ are constants. The first term of Eq. (2) represents the contribution from interfacial polarization, and the second term represents the contribution from bulk. The value of x is the proportion of the interfacial contribution. The solid curves in Fig. 1 represent the fitting results using Eq. (2). It can be seen that the model perfectly fits the experiment data. The interfacial contribution is significantly enhanced from 1.0% at 1 MHz to 83.6% at 10 kHz for sample A, and from 0.1% at 1 MHz to 92.1% at 1 kHz for sample B. Thus, the dielectric tunability is significantly enhanced by the interfacial polarization at low frequencies. At high frequencies, the tunability is greatly suppressed by the domination of bulk contribution. This enhancement further convinces that the giant tunablity is induced by the interfacial polarization.

### B. Negative dielectric tuanbility

Figure 4 shows the frequency dependence of dielectric tunability ($n_r$-F curves) measured using Au electrodes at 10 V. It is interesting that a valley occurs in the $n_r$-F curves. As shown in Fig. 4(a), the tunability of sample A decreases first slowly with the increase of frequency, and then drops rapidly to a minimum of about 7% at frequency labeled $f_{min}$ at 3.2x$10^5$ Hz. As the frequency is further increased, the tunabiliy increases to 23% at 1 MHz. A similar valley with a region of negative tunability was observed in sample B. The maximum negative tunability is about -17% at $f_{min}$=7.5x$10^4$ Hz. The values of $f_{min}$ are close to the values that separate the interfacial and bulk responses in the impedance spectra (3.0x$10^5$ Hz for sample A; 5.5x$10^4$ Hz for sample B). The $n_r$-F curves can be divided into three regions: low frequency region (LFR), bottom region (BR), and high frequency region (HFR), respectively. The LFR is dominated by the interfacial response, while the HFR is dominated by the bulk response. Therefore, the BR is a transition region, in which the interfacial and bulk responses coexist. The valley is certainly caused by the transition from interfacial response to bulk response with the increase of frequency. The inset of Fig. 4(b) shows the $N_c$-$E$ curves of sample B at 1 kHz and 100 kHz, respectively. The two $N_c$-$E$ curves exhibit opposite varying trend: with the increase of DC field, the 1 kHz curve is dropped rapidly at first and then followed by a slow decrease, while a slow increase followed by a rapid enhancement is observed in the 100 kHz $N_c$-$E$ curve.

Table I Fitting parameters of Fig. 5 using the equivalent circuit shown in Fig. 5(b), where $R_i$ and $C_i$ describe the interfacial response, $R_b$ and $C_b$ describe the bulk response, and V is the DC voltage.

| Sample   | V (V) | $R_i$ (Ω) | $C_i$ (nF) | $R_b$ (Ω) | $C_b$ (nF) |
|----------|-------|-----------|------------|-----------|------------|
| A        | 0     | 70        | 180        | 60        | 0.45       |
| (x=0.07) | 10    | 10        | 120        | 60        | 0.30       |
| B        | 0     | 800       | 250        | 400       | 0.15       |
| (x=0.48) | 10    | 55        | 110        | 390       | 0.15       |

The negative tunability can be explained based on an equivalent circuit model with two parallel RC elements in series, which is usually used to describe the Maxwell-Wagner relaxation.[16] Figure 5 shows the dielectric spectra near the transition region when Au electrodes were used. The solid curves are the fitting results with the equivalent circuit in Fig. 5(b). The parameters are listed in Table I. The characteristic frequency (peak frequency of the imaginary dielectric spectra) of the relaxation can be expressed as[17]

$$f_p = \frac{1}{2\pi\tau} = \frac{R_b + R_i}{2\pi R_b R_i (C_b + C_i)}, \quad (3)$$

where, $R_b$ and $R_i$ refer to bulk and interfacial resistance, $C_b$ and $C_i$ refer to bulk and interfacial capacitance. For both samples, $C_b$ is much smaller than $C_i$ and $R_b$ is comparable to $R_i$, so Eq. (3) can be approximately expressed as

$$f_p \approx \frac{1}{2\pi R_i C_i} + \frac{1}{2\pi R_b C_i}. \quad (4)$$

With the increase of DC bias voltage, both $C_i$ and $R_i$ are reduced significantly and $R_b$ is nearly invariable. Therefore, $f_p$ will be increased, that results in a shift of the step in dielectric spectra toward higher frequencies. The bulk capacitance of sample A has a considerable reduction (23% at 1 MHz), as shown in Fig. 5(a), so the curve measured at 10 V is still below the curve measured at 0 V, and a positive tunablity is the result. However, as shown in Fig. 5(b), the bulk capacitance of sample B is almost invariable with the increase of DC bias voltage. Therefore, part of the curve measured at 10 V is above the curve measured at 0 V, leading to the negative tunability. Based on above analyses, we suggest that the appearance of the negative tunability probably needs two conditions: one is the shift of voltage dependence of dielectric step caused by the interfacial polarization, the other one is small (close to zero) bulk tunability. The physical mechanism of the negative tunability in sample B is explained below. In the LFR, the interfacial contribution dominates. The applied DC electric filed widens the depletion layer and reduces greatly the capacitance, resulting in a giant positive tuanbility. In the HFR, the bulk contribution dominates, and the bulk capacitance will not be changed significantly with such a small electric filed, resulting in a very small (close to zero) tunability. In the BR, both the interfacial and bulk contributions have considerable effect on the capacitance. The applied DC electric field may have two effects on the interfacial polarization: one is the enhancement of the broadness of the depletion layer, the other one is the enhancement of the interfacial contribution with more charge carriers accumulated at the interface of electrode/sample. Though the former effect reduces the capacitance, the latter increases the capacitance. In the BR, these two effects coexist, and compete with each other. The competition result is that the latter dominates in the BR region and the negative tunablity emerges.

J. Sigman *et al* observed the negative tunability in $KNbO_3/KTaO_3$ superlattices.[18-19] They suggested that the negative tunability was caused by a ferroelectric-antiferroelectric phase transition with the temperature varying. Recently, X. Hu *el al* reported that the negative tunablity was detected in $0.9Pb(Fe_{1/2}Nb_{1/2})O_3/0.1CaTiO_3$ ceramics in lower DC bias field.[20] In our case, the experiment was performed at room temperature, so there was no phase transition. Moreover, no matter how lower the DC bias field was, the negative tunablity could not be observed when the experimental frequency was lower than 10 kHz for sample B. The negative tunbility is only dependent on frequency, indicating that the interfacial polarization is the main origin of the negative tunablity.

**Conclusions**

The dielectric tunability properties of PFNT single crystals were investigated in the low electric field at room temperature. The giant and negative dielectric tunability was detected. The complex impedance analysis and the experiment using different electrode materials suggest that this giant tunability might be mainly caused by the interfacial polarization at the interface of electrode/sample. The negative tunability which was only observed in sample B in the transition region from interfacial to bulk response can be also attributed to the interfacial polarization.

**Acknowledgements**

This work is supported by the National Natural Science Foundation of China (10401004). Professor F. Lu and Doctor Q. J. Cai are gratefully acknowledged for the dielectric measurement. The authors also thank Prof. H. Luo of Shanghai Institute of Ceramics (CAS) for his providing us part of samples.

**Figure captions**

Figure 1 (colour online) Normalized capacitance as a function of DC bias field for sample A (a) and sample B (b) at different frequencies using Au electrodes.

Figure 2 (colour online) Impedance spectra under several DC bias voltages for sample A (a) and sample B (b) using Au electrodes. The insets are frequency dependent dielectric spectra under these DC bias voltages.

Figure 3 (colour online) Normalized capacitance as a function of DC bias field for sample A and sample B using Al electrodes.

Figure 4 (colour online) Frequency dependence of tunability of sample A (a) and sample B (b) using Au electrodes. The inset of (b) is normalized capacitance of sample B as a function of DC bias field at 1k and 100k Hz.

Figure 5 (colour online) Dielectric spectra near the transition region from extrinsic to intrinsic response for sample A (a) and sample B (b) when Au electrodes are used.

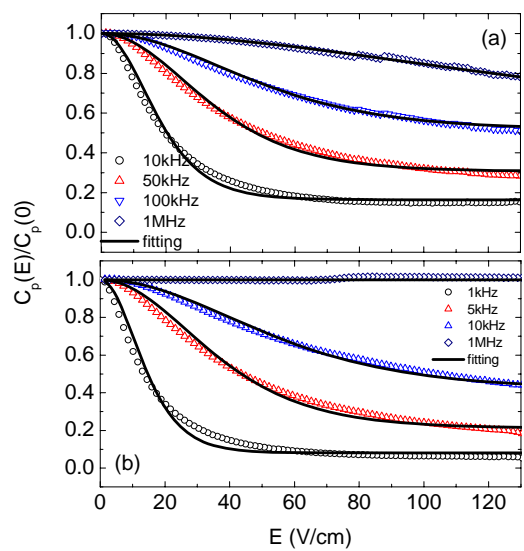

Figure 1

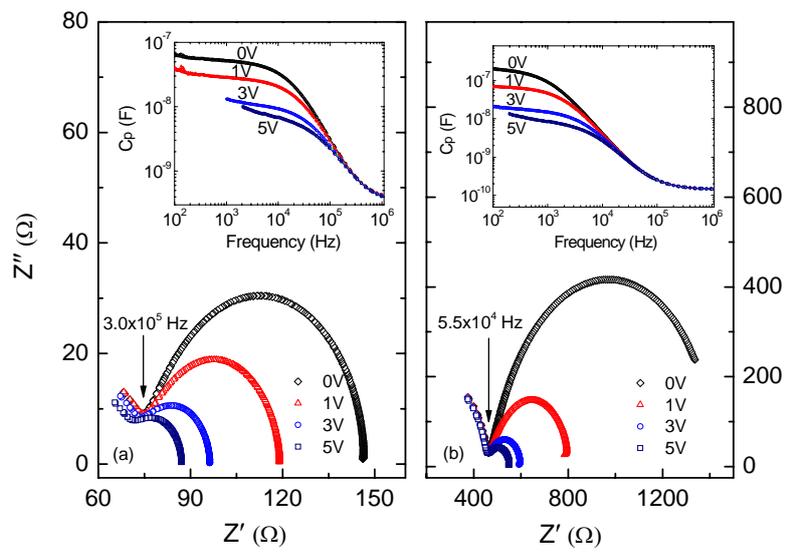

Figure 2

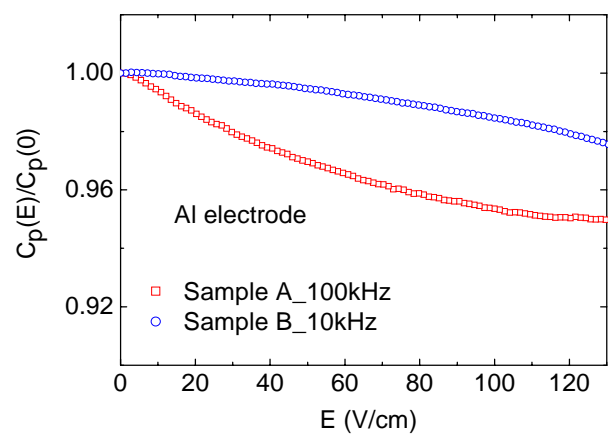

Figure 3

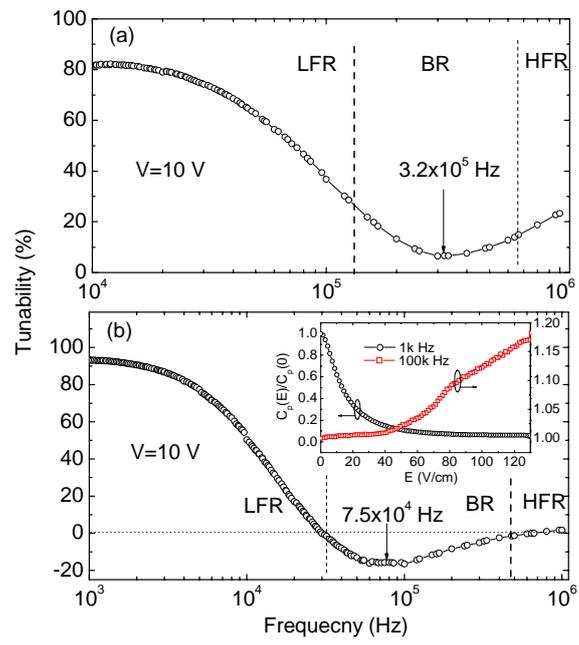

Figure 4

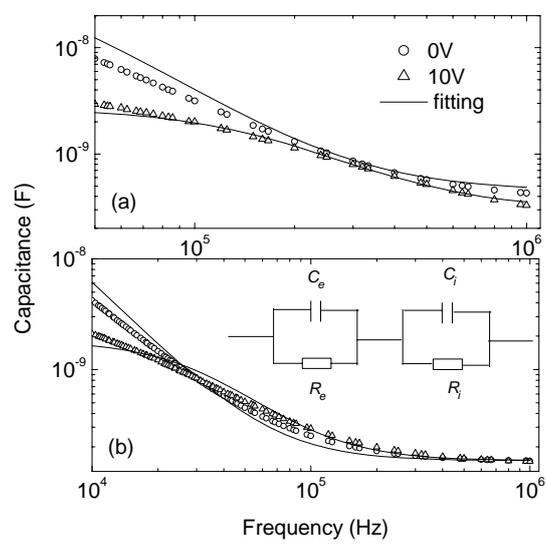

Figure 5